\documentclass{article}

\usepackage{example/spconf}
\usepackage{amsmath,amssymb,graphicx,hyperref}
\usepackage{float,arydshln}
\hypersetup{hidelinks}

\title{CONTINUAL RESIDUAL MEMORY FOR GRADIENT-FREE TEST-TIME ADAPTATION IN SPEECH ENHANCEMENT}

\name{
Yijia Shan$^{1,3}$,
Tianrui Wang$^{4}$,
Zixiang Wang$^{6}$,
Sheng Wang$^{2,7}$,
Yao Li$^{7}$,
Yu Wang$^{5}$,
Xie Chen$^{1}$
}

\address{
$^{1}$X-LANCE Lab, Shanghai Jiao Tong University, China
\quad
$^{2}$Shanghai Jiao Tong University, China\\
$^{3}$SJTU Paris Elite Institute of Technology, Shanghai Jiao Tong University, China\\
$^{4}$Tianjin University, China
\quad
$^{5}$School of Artificial Intelligence, Shanghai Jiao Tong University, China\\
$^{6}$China University of Mining and Technology, China
\quad
$^{7}$Shanghai Aviation Electric Co., Ltd., China
}

\begin{document}

\ninept
\maketitle

\begin{abstract}
Offline-trained speech enhancement models can degrade under deployment
mismatch. Test-time adaptation (TTA) addresses this problem using unlabeled
test observations, but existing approaches mainly rely on online parameter
optimization and do not explicitly exploit the residual discarded by the
enhancement backbone. We propose Continual Residual Memory (CRM), a
gradient-free continual TTA framework for selective residual recovery. CRM first estimates bounded time-frequency residual recovery from the current
utterance alone and then uses causal prototype memory from preceding utterances
to refine this estimate. During deployment, adaptation is driven solely by the evolving
memory state while the backbone and CRM modules remain frozen. Across
DNS, EARS-D, and Libri-MUSAN, CRM improves all seven reported metrics over
frozen CMGAN and achieves the best PESQ, CSIG, and COVL on all three test sets
among the compared methods. CRM remains lightweight and shows cross-backbone applicability on five
additional backbones. Code is available at
\url{https://github.com/Ada312/continual-residual-memory}.
\end{abstract}

\section{Introduction}
\label{sec:intro}

Speech enhancement (SE) aims to recover target speech from noisy or degraded observations. Modern neural enhancers are typically trained offline and frozen at deployment, but mismatches in speakers, noise types, or acoustic conditions can degrade performance. Test-time adaptation (TTA) addresses this mismatch by using unlabeled test observations to adapt a pretrained model without requiring target-domain clean speech. Existing SE TTA and related methods include self-supervised test-time training \cite{behera2025tttse}, embedding-based adaptation such as LaDen \cite{raichle2026laden}, mask polarization such as MPol \cite{raichle2026mpol}, continual self-training \cite{tzinis2022remixit}, and knowledge-distillation-based personalization \cite{kim2024zeroshot}. These approaches generally rely on parameter updates during adaptation and may depend on backbone-specific internal representations. However, enhancement itself provides an observable test-time signal: the discarded residual, i.e., the difference between the noisy input and the enhanced output. This residual contains what the backbone suppresses and, under mismatch, may include useful speech lost through over-suppression. Existing SE TTA methods do not explicitly exploit this residual for adaptation.

Previous studies show that reusing discarded information can reduce processing distortion \cite{iwamoto2022artifacts,ochiai2024processing}, while residual remixing and post-processing can alleviate over-suppression \cite{qu2026rethinkrefine,zhou2025postprocessing}. These studies show that residual information can be useful, but exploiting it requires selective recovery because directly restoring the full residual would also reintroduce suppressed noise. Moreover, prior methods do not use preceding residual history to refine subsequent recovery decisions in a continual TTA process.

In this work, we propose Continual Residual Memory (CRM), a gradient-free continual TTA
framework that combines utterance-local residual recovery with causal prototype
memory. CRM first estimates bounded time-frequency (TF) residual recovery from the current utterance, then refines it using residual patterns from preceding utterances.
The backbone and CRM modules remain frozen while only memory evolves. Our contributions are: (1) TF-wise selective residual recovery for a frozen enhancement backbone; (2) causal prototype memory that uses preceding residual patterns to refine utterance-local residual recovery while preserving the no-memory estimate as a bounded fallback; (3) consistent
improvements across multiple target test sets and both generative and
discriminative enhancement backbones with modest overhead.

\section{Method}
\label{sec:method}

\begin{figure*}[t]
\centering
\includegraphics[width=\textwidth]{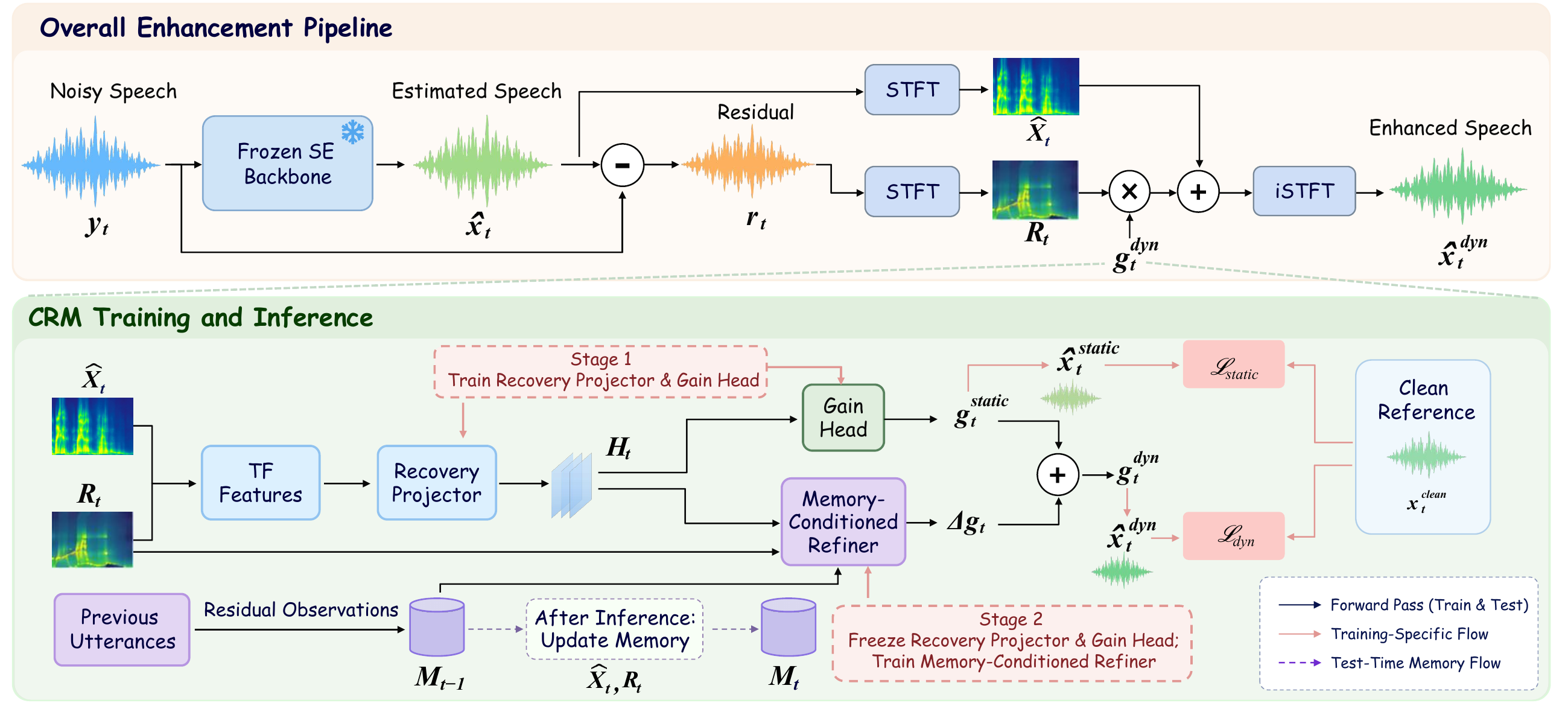}
\vspace{-5mm}
\caption{CRM overview. The upper panel shows the overall enhancement pipeline with residual recovery around a frozen backbone, while the lower panel shows two-stage training and causal memory flow. At test time, only memory is updated after each utterance.}
\label{fig:overview}
\end{figure*}

CRM augments a frozen enhancer with utterance-local residual recovery and
memory-conditioned refinement. At test time, only memory evolves through the
read--infer--write procedure in Fig.~\ref{fig:overview}.

\vspace{-2mm}

\subsection{Backbone and Residual Representation}

For utterance $t$, let $y_t$ denote the noisy waveform and
$\hat{x}_t=G_\theta(y_t)$ the output of a frozen enhancement backbone
$G_\theta$. The discarded residual is
$r_t=y_t-\hat{x}_t$, which contains both removed noise and potentially useful
speech lost through over-suppression. Restoring it unconditionally would reintroduce noise. CRM therefore performs selective recovery in the TF domain, allowing TF bins to receive different recovery gains. We
define $\hat{X}_t=\operatorname{STFT}(\hat{x}_t)$ and
$R_t=\operatorname{STFT}(r_t)$. Using a power-compression exponent $p$, we
construct seven TF feature maps from the backbone estimate and residual:
power-compressed real and imaginary components of both spectra, a stabilized
log-magnitude ratio, and sine/cosine phase-difference maps. These features
capture spectral content, relative residual strength, and local phase
agreement, and are concatenated as $\Phi_t$ for the Recovery Projector.

\subsection{Utterance-Local Residual Recovery}

Utterance-local residual recovery uses only the current backbone estimate and
discarded residual to predict a bounded recovery gain at each TF bin, serving
as CRM's no-memory fallback.

As shown in Fig.~\ref{fig:overview}, the seven-channel TF feature $\Phi_t$ constructed from $\hat{X}_t$ and $R_t$ is fed to the Recovery Projector $P_s$. The projector lifts the input to 32 channels and applies six gated dilated residual blocks, producing the hidden representation $H_t=P_s(\Phi_t)$. The Gain Head then predicts the TF logit map $\ell_t$ with a $1\times1$ convolution, yielding
\begin{equation}
\begin{aligned}
g_t^{\mathrm{static}}
&=g_{\max}\operatorname{sigmoid}(\ell_t),\\
\hat{X}_t^{\mathrm{static}}
&=\hat{X}_t+g_t^{\mathrm{static}}\odot R_t ,
\end{aligned}
\label{eq:static}
\end{equation}
where $\odot$ denotes pointwise multiplication over TF bins, and
$g_{\max}$ bounds the maximum recovery.
Here, $g_t^{\mathrm{static}}$ is a real-valued TF-wise recovery gain that controls how much of the discarded residual is restored at each TF bin. This parameterization constrains the local correction to a bounded portion of the residual already discarded by the backbone. This step uses only current-utterance evidence, while $H_t$ is retained for subsequent memory-conditioned refinement.

Offline training with clean references minimizes
\begin{equation}
\begin{aligned}
\mathcal L_{\mathrm{static}}
={}&\mathcal L_{\mathrm{time}}
+\lambda_{\mathrm{complex}}\mathcal L_{\mathrm{complex}}
+\lambda_{\log}\mathcal L_{\log\mathrm{mag}}\\
&+\lambda_{\mathrm{proj}}\mathcal L_{\mathrm{proj}}
+\lambda_{\mathrm{SI}}\mathcal L_{\mathrm{SI\mbox{-}SDR}}
+\lambda_{\mathrm{trust}}\mathcal L_{\mathrm{trust}} .
\end{aligned}
\label{eq:static_loss}
\end{equation}
The six terms respectively denote normalized waveform $L_1$,
complex-spectrum $L_1$, log-magnitude reconstruction, gain projection,
negative zero-mean SI-SDR, and trust regularization against excessive recovery,
with the $\lambda$ terms as their weights. Let
$x_t^{\mathrm{clean}}$ denote the clean training reference and
$X_t^{\mathrm{clean}}=\operatorname{STFT}(x_t^{\mathrm{clean}})$. The oracle TF recovery gain is obtained by projecting the desired clean correction $X_t^{\mathrm{clean}}-\hat X_t$ onto the discarded residual:
\begin{equation}
g_t^\star=
\Pi_{[0,g_{\max}]}
\left(
\frac{\Re\{R_t^\ast(X_t^{\mathrm{clean}}-\hat X_t)\}}
{\max(|R_t|^2,\epsilon_{\mathrm{proj}})}
\right).
\label{eq:oracle_gain}
\end{equation}
Here $\Re\{\cdot\}$ and $(\cdot)^\ast$ denote real part and complex conjugation, $\Pi_{[0,g_{\max}]}$ denotes projection onto the valid recovery-gain interval, and $\epsilon_{\mathrm{proj}}>0$ is a stabilizing constant.
The gain projection term \(\mathcal L_{\mathrm{proj}}\) is a residual-power-weighted Smooth-\(L_1\) loss between predicted and oracle gains.

\subsection{Causal Prototype Memory and Refinement}

Utterance-local recovery provides a no-memory estimate but cannot exploit residual patterns that recur across successive test utterances. CRM therefore uses causal prototype memory to summarize prior residual history and refine this estimate, allowing the recovery gain to increase or decrease according to memory-conditioned evidence.

\noindent\textbf{Prototype Memory.}
Before utterance $t$, $M_{t-1}$ contains at most
$K_{\max}$ residual log-power prototypes with frequency-wise mean $\mu_m$,
variance $v_m$, and observation count $n_m$, where $m$ indexes prototypes. The current residual is summarized as
\begin{equation}
u_t(f) =
\operatorname{median}_{\tau}
\log\left(
\max\left(|R_t(f,\tau)|^2,\epsilon_{\mathrm{mem}}\right)
\right),
\end{equation}
where $f$ and $\tau$ denote frequency bins and frames, and
$\epsilon_{\mathrm{mem}} = 10^{-8}$ is used for numerical stability.
Let $\bar{\mu}_m$ and $\bar{u}_t$ denote frequency means, with
$\tilde{\mu}_m=\mu_m-\bar{\mu}_m$ and
$\tilde{u}_t=u_t-\bar{u}_t$ denoting mean-centered spectral profiles.

CRM matches the current summary $u_t$ to each prototype using

\begin{equation}
D_m(u_t)=
\operatorname{mean}_f
\left[
\frac{(\tilde{\mu}_m-\tilde{u}_t)^2}
{v_m+\lambda_{\mathrm{var}}}
\right]
+\lambda_{\mathrm{level}}
(\bar{\mu}_m-\bar{u}_t)^2 ,
\label{eq:memory_distance}
\end{equation}
where $\lambda_{\mathrm{var}}>0$ stabilizes variance normalization and
$\lambda_{\mathrm{level}}$ controls overall log-power mismatch.

The distances are converted to normalized weights with temperature \(\tau_{\mathrm{mem}}\) by \(p_m\propto\exp(-D_m/\tau_{\mathrm{mem}})\), and CRM sorts prototypes by decreasing \(p_m\) and retains the smallest subset whose cumulative weight reaches the readout threshold \(\gamma_{\mathrm{read}}\).

The retained weights $p_m$ are renormalized and used to aggregate the prototype
means $\mu_m$ and variances $v_m$. CRM then constructs six context maps: a
frequency-centered aggregate prototype profile from $\sum_m p_m\mu_m$; the
normalized deviation of current residual log-power $\log |R_t|^2$ from the
aggregate mean; stability from the aggregated variance; history confidence
from prototype counts $n_m$; query similarity $\exp(-\min_m D_m(u_t))$; and
the normalized entropy of $p_m$. Their stack forms $Z_t$, with $Z_t=0$ for
empty memory.

\noindent\textbf{Memory-Conditioned Refiner.}
The refiner uses two lightweight convolutional branches to model current-utterance evidence and memory context separately. The basis branch transforms
$H_t$ through a 32-channel convolutional projection, a depthwise convolution,
and an output projection to produce $d$ TF basis maps
$B_t=\{b_{t,k}\}_{k=1}^{d}$, where $d$ is the low-rank dimension. These maps
represent candidate recovery patterns supported by the current utterance. The
coordinate branch applies an analogous convolutional mapping to $Z_t$ and
produces $d$ coefficient maps $C_t=\{c_{t,k}\}_{k=1}^{d}$. These coefficients
determine how strongly memory modulates the corresponding basis patterns.
This factorization lets memory refine a compact set of
current-utterance recovery patterns rather than predict a full correction
directly.
A pointwise low-rank interaction combines the two branches into a signed correction $\Delta g_t$, which is added to $g_t^{\mathrm{static}}$ and projected onto the valid gain range to obtain $g_t^{\mathrm{dyn}}$:
\begin{equation}
\begin{aligned}
\Delta g_t &=
\alpha_{\mathrm{dyn}}\rho_t
\tanh\!\left(
\frac{1}{\sqrt d}
\sum_{k=1}^{d} b_{t,k}\odot c_{t,k}
\right),\\
g_t^{\mathrm{dyn}}
&=
\Pi_{[0,g_{\max}]}
\left(g_t^{\mathrm{static}}+\Delta g_t\right),\\
\hat X_t^{\mathrm{dyn}}
&=
\hat X_t+g_t^{\mathrm{dyn}}\odot R_t .
\end{aligned}
\label{eq:dynamic}
\end{equation}

The final enhanced waveform $\hat{x}^{\mathrm{dyn}}_t$ is obtained by inverse STFT of
$\hat{X}^{\mathrm{dyn}}_t$. Here $\alpha_{\mathrm{dyn}}$ controls the correction magnitude, and
$\rho_t=\mathbb{I}[M_{t-1}\neq\varnothing]$ activates the correction only when
memory is available, where $\mathbb{I}[\cdot]$ is the indicator function. Positive
$\Delta g_t$ increases recovery, while negative $\Delta g_t$ decreases it.
For empty memory, $\rho_t=\Delta g_t=0$, so CRM reduces to the no-memory
estimate.

\noindent\textbf{Training and Causal Memory Update.}
As shown in Fig.~\ref{fig:overview}, training proceeds in two stages: Stage 1 optimizes the Recovery Projector and Gain Head with $\mathcal L_{\mathrm{static}}$; Stage 2 freezes them and optimizes the basis and coordinate branches within the Memory-Conditioned Refiner with \(\mathcal L_{\mathrm{dyn}}\). The base objective \(\mathcal L_{\mathrm{base}}\) combines the reconstruction and projection terms defined above with SI-SDR, segmental-SNR, trust, and differentiable PESQ losses.
To guard against unreliable history, training uses matched, shuffled, and empty-memory conditions. Matched memory uses the corresponding causal
history, shuffled memory uses histories from other training examples, and empty memory disables historical context, reducing CRM to the no-memory estimate.

Using the weighted sum of the waveform, complex-spectrum, and log-magnitude
reconstruction terms defined above, $\mathcal R_{\mathrm{match}}$ and
$\mathcal R_{\mathrm{shuf}}$ penalize reconstruction degradation relative to
the empty-memory fallback under matched and shuffled history, respectively.
A shuffled-history consistency term \(\mathcal C_{\mathrm{shuf}}\) further limits waveform drift under harmful shuffled history, with a soft gate of temperature \(\tau_{\mathrm{cf}}\) applied within \(\mathcal C_{\mathrm{shuf}}\) to increase the penalty when shuffled history degrades reconstruction. The refinement objective is
\begin{equation}
\mathcal L_{\mathrm{dyn}}
=
\mathcal L_{\mathrm{base}}
+
\lambda_{\mathrm{cf}}
\left(
\mathcal R_{\mathrm{match}}
+
\mathcal R_{\mathrm{shuf}}
+
\lambda_{\mathrm{cons}}\mathcal C_{\mathrm{shuf}}
\right).
\label{eq:dynamic_loss}
\end{equation}
Here $\lambda_{\mathrm{cf}}$ controls the overall counterfactual regularization
and $\lambda_{\mathrm{cons}}$ weights the shuffled-history consistency term.
The objective therefore trains CRM to exploit useful history while protecting
the no-memory fallback when memory is unreliable.

At deployment, CRM follows a causal read--infer--write procedure initialized
with $M_0=\varnothing$. Utterance $t$ is enhanced using only $M_{t-1}$, and
its residual is written to memory only after the enhanced output has been
produced. For memory writing, we measure the fraction of frame energy contributed by
the backbone output:
\begin{equation}
q_{t,\tau}=
\frac{\sum_f |\hat X_t(f,\tau)|^2}
{\sum_f |\hat X_t(f,\tau)|^2+
 \sum_f |R_t(f,\tau)|^2+
 \epsilon_{\mathrm{mem}}},
\label{eq:backbone_occupancy}
\end{equation}
where lower values indicate more residual-dominant frames. Using Eq.~\eqref{eq:backbone_occupancy}, we select the 30\% lowest-occupancy frames and define \(o_t\) as their median residual log-power profile.

Unlike the full-utterance query $u_t$ used for memory retrieval, $o_t$ focuses
memory updates on residual-dominant structure. We evaluate \(D_m(o_t)\) using the same prototype distance in Eq.~\eqref{eq:memory_distance} and match \(o_t\) to the nearest prototype. If sufficiently novel, \(o_t\) may create a new prototype; otherwise, it updates the matched prototype statistics \((\mu_m,v_m,n_m)\). Redundant prototypes
may be merged, and when memory is full, only stale prototypes are eligible for
replacement. This write operation produces $M_t$ only after inference, so
utterance $t$ is enhanced using $M_{t-1}$ and cannot influence its own output.

\section{Experimental Setup}
\label{sec:setup}

\noindent\textbf{Datasets.} CRM is trained only on EARS-WHAM, constructed from
EARS \cite{richter2024ears} and WHAM! \cite{wichern2019wham}. We use 8,192 training mixtures and a source-domain held-out set of 632 speaker-disjoint mixtures for checkpoint and hyperparameter selection. Target-domain clean references are used only for evaluation. The evaluation uses DNS 2020 synthetic no-reverberation (150 utterances) \cite{reddy2020dns}, EARS-D (886 mixtures), constructed from the EARS test split and DEMAND noise at SNRs uniformly sampled from $[-2.5,17.5]$~dB \cite{raichle2026laden,richter2024ears,thiemann2013demand}, and Libri-MUSAN (2,620 mixtures), constructed from LibriSpeech test-clean and MUSAN music at $\{-5,0,5,10\}$~dB \cite{panayotov2015librispeech,snyder2015musan}. Each dataset is processed independently as a fixed-order causal stream
initialized with empty memory.

\noindent\textbf{Training.}
Audio is sampled at 16~kHz using a 512-point Hann STFT and a 128-sample hop.
Both stages use 2-s crops, batch size 8, AdamW optimizer with weight decay
$10^{-4}$, and cosine scheduling. The Recovery Projector and Gain Head are trained for four epochs with a learning rate of \(3\times10^{-4}\) and then frozen. The Memory-Conditioned Refiner is trained for three epochs with a
learning rate of $10^{-3}$ using causal EARS-WHAM histories grouped by WHAM
recording location. We use $K_{\mathrm{train}}=2$ and memory dropout 0.20 to
expose the refinement module to missing historical context, with
$(\tau_{\mathrm{cf}},\lambda_{\mathrm{cf}},\lambda_{\mathrm{cons}})
=(0.02,0.10,0.10)$. During refinement training, memory readout uses
$\tau_{\mathrm{mem}}=0.50$ without posterior-mass truncation. At deployment,
we retain $\tau_{\mathrm{mem}}=0.50$, use $K_{\max}=64$ as the memory capacity,
and set $\gamma_{\mathrm{read}}=0.95$ to select the highest-weight prototype
subset. We set
$(p,d,g_{\max},\alpha_{\mathrm{dyn}})=(0.30,4,0.10,0.08)$.
Memory updates use a novelty threshold of 0.35 for prototype creation, a merge threshold of 0.05, and a stale age of 20 utterances for replacement.

\noindent\textbf{Baselines and Evaluation.} The main backbone is
CMGAN \cite{cao2022cmgan} trained on EARS-WHAM. We compare CRM with LaDen
\cite{raichle2026laden} and MPol \cite{raichle2026mpol}, two representative
gradient-based TTA methods for speech enhancement. For controlled comparison, LaDen, MPol, and CRM all start from the same CMGAN checkpoint released with the LaDen/MPol implementation and are evaluated under the same test protocol. LaDen and MPol
adapt selected CMGAN parameters during testing. For CRM, both CMGAN and the CRM
modules remain frozen and adaptation occurs only through the prototype-memory
state. We report PESQ \cite{rix2001pesq}, STOI
\cite{taal2011stoi}, CSIG/CBAK/COVL \cite{hu2008objective}, clipped segmental
SNR (SSNR), and SI-SDR \cite{leroux2019sisdr}.
\section{Results and Analysis}
\label{sec:results}

\subsection{Main Comparison}

To evaluate CRM under mismatched test conditions, Table~\ref{tab:main}
compares it with LaDen and MPol on the three target test sets. CRM improves all seven reported metrics over the frozen backbone,
with PESQ gains of 0.11, 0.02, and 0.11 on DNS, EARS-D, and Libri-MUSAN,
respectively. It also achieves the best PESQ, CSIG, and COVL across all three
datasets. These results demonstrate CRM's effectiveness for cross-domain
enhancement.

\begin{table}[!ht]
\caption{Comparison of test-time adaptation methods on three target test sets, all using the same CMGAN backbone checkpoint.}
\label{tab:main}
\vspace{1pt}
\centering
\setlength{\tabcolsep}{1.0pt}
\begin{tabular}{@{}l@{\hspace{1pt}}ccccccc@{}}
\hline
Method &
PESQ$\uparrow$ &
STOI$\uparrow$ &
CSIG$\uparrow$ &
CBAK$\uparrow$ &
COVL$\uparrow$ &
SSNR$\uparrow$ &
SI-SDR$\uparrow$ \\
\hline

\multicolumn{8}{c}{\textit{DNS}} \\
\hdashline
CMGAN
& 3.00 & 0.967 & 4.19 & 3.47 & 3.60 & 9.00 & 15.08 \\
\hspace{0.6em}+ LaDen
& 3.00 & 0.967 & 4.21 & 3.47 & 3.61 & 9.06 & 15.06 \\
\hspace{0.6em}+ MPol
& 3.00 & \textbf{0.968} & 4.22 & 3.48 & 3.62 & 9.10 & 15.11 \\
\hspace{0.6em}+ CRM
& \textbf{3.11} & \textbf{0.968} & \textbf{4.36}
& \textbf{3.56} & \textbf{3.75} & \textbf{9.23}
& \textbf{15.34} \\

\hline
\multicolumn{8}{c}{\textit{EARS-D}} \\
\hdashline
CMGAN
& 2.62 & 0.846 & 3.76 & 2.82 & 3.17 & 2.41 & 5.06 \\
\hspace{0.6em}+ LaDen
& 2.57 & \textbf{0.849} & 3.83 & \textbf{2.88}
& 3.18 & \textbf{3.59} & \textbf{5.64} \\
\hspace{0.6em}+ MPol
& 2.61 & 0.848 & 3.81 & 2.82 & 3.19 & 2.43 & 5.04 \\
\hspace{0.6em}+ CRM
& \textbf{2.64} & \textbf{0.849} & \textbf{3.90}
& 2.86 & \textbf{3.26} & 2.63 & 5.20 \\

\hline
\multicolumn{8}{c}{\textit{Libri-MUSAN}} \\
\hdashline
CMGAN
& 2.16 & 0.903 & 3.40 & 2.62 & 2.74 & 3.83 & 9.64 \\
\hspace{0.6em}+ LaDen
& 2.16 & 0.903 & 3.54 & 2.67 & 2.81 & \textbf{4.42} & 9.61 \\
\hspace{0.6em}+ MPol
& 2.18 & \textbf{0.905} & 3.46 & 2.65 & 2.78 & 3.92 & \textbf{9.87} \\
\hspace{0.6em}+ CRM
& \textbf{2.27} & 0.904 & \textbf{3.59}
& \textbf{2.73} & \textbf{2.90} & 4.07 & 9.83 \\
\hline

\end{tabular}
\end{table}
\vspace{-1mm}

\subsection{Component Analysis of CRM}

To analyze the complementary roles of the two steps in CRM's recovery process, Table~\ref{tab:memory_effect} decomposes the full framework into utterance-local residual recovery and memory-conditioned refinement. On DNS, full CRM improves PESQ from 3.00 to 3.11: local recovery reaches 3.07, and memory further raises it to 3.11. Memory provides additional PESQ gains on all three datasets. On DNS, the
$+0.04$ gain has a 95\% paired utterance-level bootstrap CI of
$[0.03,\,0.04]$, based on 20,000 resamples.

\begin{table}[!ht]
\caption{Component-wise analysis of CRM in terms of PESQ ($\uparrow$).}
\label{tab:memory_effect}
\vspace{1pt}
\centering
\setlength{\tabcolsep}{4.5pt}
\begin{tabular}{@{}lcc@{\hspace{8pt}}c@{}}
\hline
Method & DNS & EARS-D & Libri-MUSAN \\
\hline
Backbone
& 3.00 & 2.62 & 2.16 \\
\quad\textit{+ Utterance-local recovery}
& 3.07 & 2.61 & 2.23 \\
\quad\textit{+ Full CRM}
& \textbf{3.11} & \textbf{2.64} & \textbf{2.27} \\
\hline
\end{tabular}
\end{table}

\begin{figure}[!ht]
\centering
\begin{minipage}[t]{0.48\columnwidth}
\centering
\includegraphics[width=\linewidth]{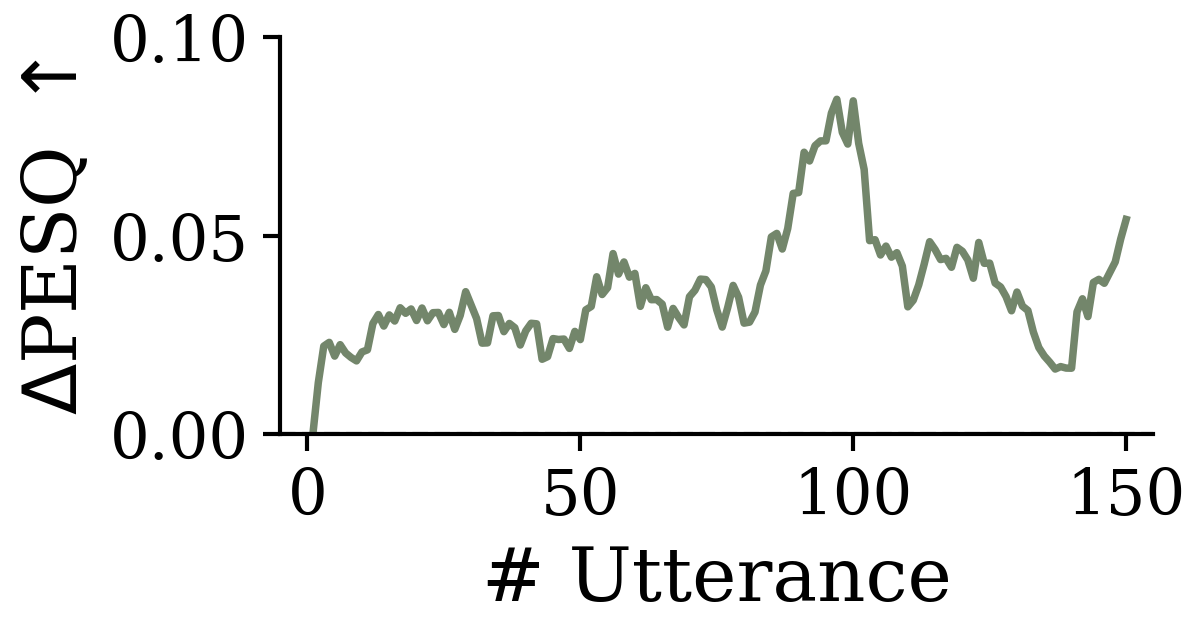}
\end{minipage}
\hfill
\begin{minipage}[t]{0.48\columnwidth}
\centering
\includegraphics[width=\linewidth]{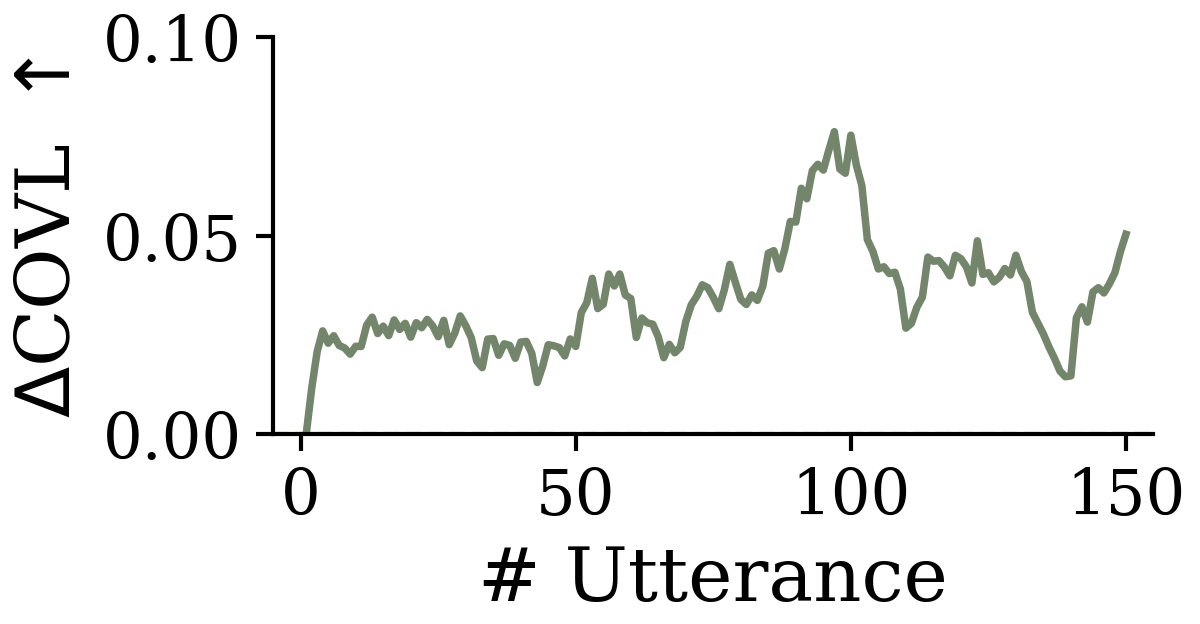}
\end{minipage}
\vspace{-3mm}
\caption{Incremental gains from CRM's memory-conditioned refinement step on the ordered DNS stream.}
\label{fig:continual_progression}
\end{figure}

Figure~\ref{fig:continual_progression} further shows the incremental gains contributed by CRM's memory-conditioned refinement step beyond utterance-local recovery, using a causal 10-utterance moving average along the ordered DNS stream. The gain at the first utterance is zero because the memory is initially empty, while the smoothed PESQ and COVL gains remain positive once historical context becomes available.

We further examine robustness to an abrupt cross-domain transition.
\textit{Matched history} uses standard causal CRM on DNS, initialized with
empty memory and updated only with preceding DNS observations.
\textit{Cross-domain history} first accumulates memory over the complete
Libri-MUSAN stream and carries it into DNS without reset. The memory is then updated causally throughout the complete DNS stream.
As shown in Table~\ref{tab:cross_domain}, cross-domain history performs comparably to matched DNS history. This indicates CRM's robustness to the abrupt domain transition, with enhancement performance preserved without resetting residual memory.

\vspace{-1mm}
\begin{table}[!ht]
\caption{Cross-domain transition analysis on DNS.}
\label{tab:cross_domain}
\vspace{1pt}
\centering
\setlength{\tabcolsep}{6pt}
\begin{tabular}{@{}lccc@{}}
\hline
History condition & PESQ $\uparrow$ & COVL $\uparrow$ & SI-SDR $\uparrow$ \\
\hline
Matched history
& 3.110 & 3.750 & 15.344 \\
Cross-domain history
& 3.111 & 3.751 & 15.345 \\
\hline
\end{tabular}
\end{table}
\vspace{-1mm}

\subsection{Cross-Backbone Applicability}

To assess cross-backbone applicability, we evaluate CRM on five frozen backbones spanning diffusion
(StoRM-50 \cite{lemercier2023storm}), flow matching
(FlowSE \cite{lee2025flowse}), discriminative
(GTCRN \cite{rong2024gtcrn}), streaming (FastEnhancer-B \cite{ahn2026fastenhancer}), and ultra-lightweight (UL-UNAS \cite{rong2026ulunas}) architectures. For each backbone, its pretrained checkpoint remains frozen, and a backbone-specific CRM is independently trained and selected using only EARS-WHAM; DNS is used solely
for evaluation. As shown in Table~\ref{tab:backbones}, CRM improves PESQ,
STOI, CSIG, COVL, and SI-SDR on all five backbones, with PESQ gains up to
0.24, supporting its applicability across diverse enhancement architectures.

\vspace{-1mm}
\begin{table}[!ht]
\caption{Cross-backbone CRM results on the DNS test set.}
\label{tab:backbones}
\vspace{3pt}
\centering
\setlength{\tabcolsep}{0pt}
\renewcommand{\arraystretch}{1.02}

\begin{tabular}{
@{}l@{\hspace{-3.25pt}}
c@{\hspace{-0.4pt}}
c@{\hspace{-0.4pt}}
c@{\hspace{-0.4pt}}
c@{\hspace{-0.4pt}}
c@{\hspace{-0.4pt}}
c@{\hspace{-0.4pt}}
c@{}
}
\hline
Model &
PESQ\kern-0.5pt$\uparrow$ &
STOI\kern-0.5pt$\uparrow$ &
CSIG\kern-0.5pt$\uparrow$ &
CBAK\kern-0.5pt$\uparrow$ &
COVL\kern-0.5pt$\uparrow$ &
SSNR\kern-0.5pt$\uparrow$ &
SI-SDR\kern-0.5pt$\uparrow$ \\
\hline

StoRM-50
& 2.69 & 0.965 & 4.09 & \textbf{3.47}
& 3.40 & \textbf{11.05} & 15.90 \\
\hspace{0.6em}\textit{+ CRM}
& \textbf{2.71} & \textbf{0.966} & \textbf{4.12}
& \textbf{3.47} & \textbf{3.42} & 11.00 & \textbf{16.08} \\

GTCRN
& 1.70 & 0.871 & 3.03 & 2.74
& 2.34 & \textbf{8.14} & 8.95 \\
\hspace{0.6em}\textit{+ CRM}
& \textbf{1.94} & \textbf{0.905} & \textbf{3.32}
& \textbf{2.83} & \textbf{2.60} & 7.88 & \textbf{9.54} \\

FastEnhancer-B
& 2.07 & 0.915 & 3.32 & 3.10
& 2.69 & \textbf{10.12} & 13.54 \\
\hspace{0.6em}\textit{+ CRM}
& \textbf{2.27} & \textbf{0.927} & \textbf{3.59}
& \textbf{3.14} & \textbf{2.92} & 9.63 & \textbf{13.74} \\

FlowSE
& 2.40 & 0.956 & 3.51 & 3.31
& 2.95 & \textbf{11.33} & 15.81 \\
\hspace{0.6em}\textit{+ CRM}
& \textbf{2.58} & \textbf{0.958} & \textbf{3.92}
& \textbf{3.41} & \textbf{3.25} & 11.14 & \textbf{15.97} \\

UL-UNAS
& 2.57 & 0.949 & 3.90 & 3.45
& 3.24 & \textbf{11.60} & 16.26 \\
\hspace{0.6em}\textit{+ CRM}
& \textbf{2.68} & \textbf{0.951} & \textbf{4.03}
& \textbf{3.49} & \textbf{3.36} & 11.35 & \textbf{16.27} \\

\hline
\end{tabular}
\end{table}
\vspace{-1mm}

\subsection{Efficiency}

CRM introduces only 37.09K parameters, corresponding to 2.02\% increase in model size and 2.20\% profiler-counted FLOPs overhead over CMGAN. 
On an RTX 4090, the real-time factor (RTF) increases slightly from 0.0340 to 0.0387 and remains well below 1.0. 
CRM therefore enables continual adaptation with
modest deployment overhead, without requiring online parameter optimization.

\section{Conclusion}
\label{sec:conclusion}

We proposed CRM, a gradient-free continual TTA framework that combines bounded
utterance-local residual recovery with causal prototype memory. Without
historical context, CRM falls back to utterance-local recovery; when history is
available, prototype memory further refines this estimate without online
parameter updates. Experiments on three target test sets show consistent
improvements over frozen CMGAN, with memory-conditioned refinement providing additional gains
beyond utterance-local residual recovery. Results on five additional backbones
support CRM's cross-backbone applicability. Overall, CRM offers a lightweight
alternative to online optimization for robust test-time speech enhancement.

\bibliographystyle{example/IEEEbib}
\bibliography{example/strings,refs}

\end{document}